%% file: sample-acmtog-SIGGRAPH-submission.tex
\documentclass[sigconf, authorversion]{acmart} 
\makeatletter
\let\@authorsaddresses\@empty
\makeatother

\usepackage{booktabs} 
\usepackage{makecell}
\usepackage{wrapfig}
\usepackage{enumitem}
\usepackage{booktabs} 

\input{symbols_format.tex}

\usepackage[ruled]{algorithm2e} 

\SetAlFnt{\small}
\SetAlCapFnt{\small}
\SetAlCapNameFnt{\small}
\SetAlCapHSkip{0pt}
\DeclareMathOperator*{\argmax}{arg\,max}

\copyrightyear{2022} 
\acmYear{2022} 
\setcopyright{acmcopyright}\acmConference[SA '22 Conference Papers]{SIGGRAPH Asia 2022 Conference Papers}{December 6--9, 2022}{Daegu, Republic of Korea}
\acmDOI{10.1145/3550469.3555390}

\begin{document}
\title{\huge Morig: Motion-aware rigging of character meshes from point clouds}

\author{Zhan Xu}
\affiliation{
\institution{UMass Amherst}
\city{Amherst}
\state{MA}
\country{USA}
}
\email{zhanxu@cs.umass.edu}

\author{Yang Zhou}
\affiliation{
\institution{Adobe Research}
\city{San Jose}
\state{CA}
\country{USA}
}
\email{yazhou@adobe.com}

\author{Li Yi}
\affiliation{
\institution{Tsinghua University}
\city{Haidian}
\state{Beijing}
\country{China}
}
\email{ericyi0124@gmail.com}

\author{Evangelos Kalogerakis}
\affiliation{
\institution{UMass Amherst}
\city{Amherst}
\state{MA}
\country{USA}
}
\email{kalo@cs.umass.edu}

\renewcommand\shortauthors{Xu, Z. et al}

\begin{abstract}
We present \emph{MoRig}, a method that automatically rigs character meshes driven by single-view point cloud streams capturing the motion of performing characters. Our method is also able to animate the 3D meshes according to the captured point cloud motion.
MoRig's neural network encodes motion cues from the point clouds into features that are informative about the articulated parts of the performing character. These motion-aware features guide the inference of an appropriate skeletal rig for the input mesh, which is then animated based on the  point cloud motion.
Our method can rig and animate diverse characters, including humanoids, quadrupeds, and toys with varying articulation. It accounts for occluded regions in the  point clouds and mismatches in the part proportions between the input mesh and captured character. Compared to other rigging approaches that ignore motion cues, MoRig produces more accurate rigs,  well-suited for re-targeting motion from captured characters.
\end{abstract}

\begin{teaserfigure}
\vspace{-3mm}
\begin{center}
 \includegraphics[width=1\textwidth]{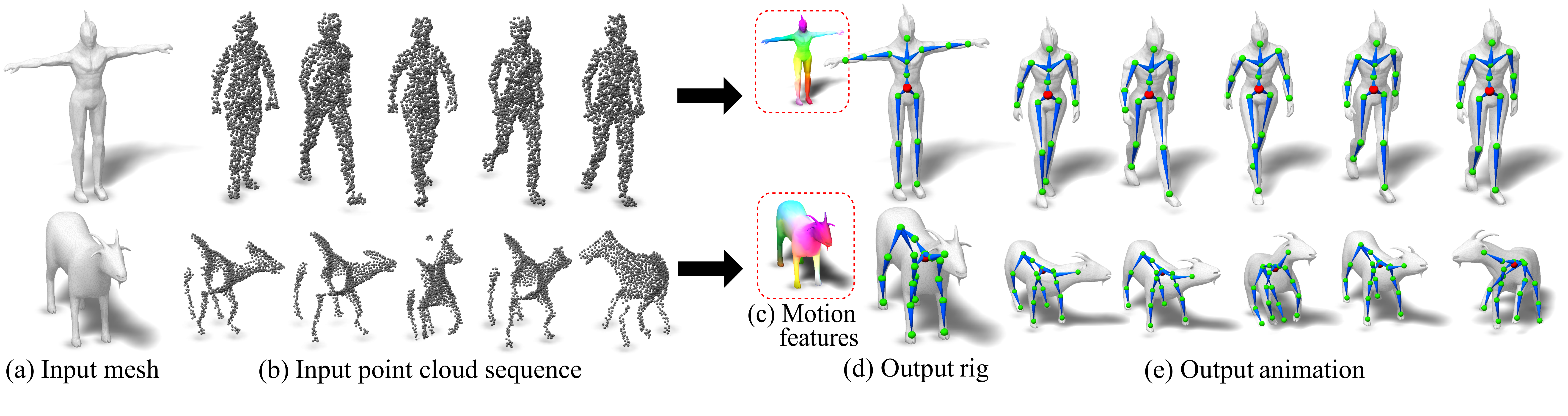}
\end{center} 
 \vspace{-6mm}
 \caption{Given (a) an input mesh and (b) a single-view point cloud sequence capturing a performing character,  our deep learning method, called \emph{MoRig}, automatically rigs and animates the mesh based on the point cloud motion. This is achieved by (c) producing motion-aware features on the mesh encoding articulated parts from the captured motion, (d) using the features to infer an appropriate skeletal rig for the mesh, and (e) re-targeting the motion from the point cloud to the rig. 
 } 
 \label{fig:teaser}
 \vspace{2mm}
\end{teaserfigure}



\keywords{character rigging, animation, neural networks, transformers\vspace{-0.5mm}}

\maketitle
\pagestyle{plain}
\input{samplebody-journals}

\appendix
\input{chapters/suppl.tex}

\end{document}

%% file: symbols_format.tex
\usepackage{xcolor}

\newcommand{\bd}{\mathbf{d}}
\newcommand{\be}{\mathbf{e}}
\newcommand{\bff}{\mathbf{f}}
\newcommand{\bg}{\mathbf{g}}
\newcommand{\bh}{\mathbf{h}}

\newcommand{\bj}{\mathbf{j}}
\newcommand{\bk}{\mathbf{k}}

\newcommand{\bq}{\mathbf{q}}

\newcommand{\bs}{\mathbf{s}}

\newcommand{\bv}{\mathbf{v}}

\newcommand{\bx}{\mathbf{x}}
\newcommand{\by}{\mathbf{y}}

\newcommand{\bA}{\mathbf{A}}

\newcommand{\bK}{\mathbf{K}}

\newcommand{\bQ}{\mathbf{Q}}

\newcommand{\bU}{\mathbf{U}}
\newcommand{\bV}{\mathbf{V}}

\newcommand{\mM}{\mathcal{M}}

\newcommand{\mR}{\mathcal{R}}

\newcommand{\mP}{\mathcal{P}}

\newcommand{\etal}{\textit{et al}.}

%% file: samplebody-journals.tex
\section{Introduction}
\label{sec:intro}
\input{chapters/introduction}

\section{Related Work}
\label{sec:related_works}
\input{chapters/related_work}

\section{Method}
\label{sec:arch}
\input{chapters/method}

\section{Training}
\label{sec:train}
\label{sec:training}
\input{chapters/training}

\section{Results}
\label{sec:exp}
\input{chapters/experiments}

\section{Discussion}
\label{sec:conclusion}
\input{chapters/conclusion}

\begin{acks}
We thank Kentaro Ko, Takuya Narihira, Tamaki Kojima for helpful discussion. We are grateful for the support from Sony Corporation.
\end{acks}

\bibliographystyle{ACM-Reference-Format}
\bibliography{sample-bibliography}


%% file: chapters/introduction.tex
With the emergence of virtual worlds, mixed reality and 3D social media, the need for diverse, high-quality, animation-ready avatars is greater than ever. To animate articulated characters, 
it is common for artists to hand-craft character rigs for meshes. These rigs are often represented as ``skeletons'' with ``skinning'' that binds the skeletons to the meshes. Unfortunately, hand-crafting skeletons and skinning  require laborious effort and extensive experience. 

To accelerate the process, a few approaches have been proposed to automate steps involved in rigging. One family of approaches fits a particular skeletal template  to the input character \cite{Baran:2007:ARA,neuralblendshape}, which can then be animated according to motion capture data for that template. 
These approaches are well-suited for motion re-targeting, yet
require different templates to be hand-crafted for characters with varying articulation structure. A different family of approaches \cite{Liu2019,rignet}
is able to infer rigs for diverse characters e.g., animals, toys, robots, and other man-made objects. However, such methods are \emph{motion-agnostic}: they process a single static mesh, and may end up creating a skeleton which is far from the users' expectations for animation. 
Our method, called \emph{MoRig}, attempts to build a common ground 
between the above two families of methods: it is able to rig diverse characters while also
considering motion cues to drive the rigging process. Specifically, given a point cloud sequence capturing the motion of a performing character, MoRig infers a skeletal rig for a mesh of a target  character. The rig matches the articulating parts of the captured character. In contrast to motion-agnostic approaches \cite{Liu2019,rignet}, MoRig is also able to animate the skeletal rig of the target character to match the captured motion. 

Inferring character rigs and  motion from point cloud guidance has several challenges. First, captured point clouds, in particular ones captured from a single view, suffer from occlusions allowing only partial observations of the moving parts in the performing character. Second, the captured character may not match well the input target mesh due to geometric differences, such as mismatches in part proportions. Third, transferring the motion from the point cloud requires controlling the produced rig to closely match this motion, while also being robust to noise and outliers. MoRig addresses the challenges
through a novel mesh deformation network that attempts to closely align the target character with each point cloud frame, while factoring out mismatches in part proportions and noise. To infer the skeletal rig, our method incorporates a neural network that considers both the geometry of the target mesh as well as
motion-aware features produced by a motion encoder module
based on transformers~\cite{vaswani2017attention}. Our method improves the rigging and skinning of diverse characters compared to state-of-the-art methods, as demonstrated  in our experiments. To summarize, our contributions are the following:

\begin{itemize}[leftmargin=*]
  \item A deformation module that aligns target meshes of articulated characters
  with point clouds under noise and occlusion.
  \item A transformer-based motion encoder that encodes motion trajectories of mesh vertices into features revealing articulated parts.  
  \item A unified pipeline of rigging, skinning, and motion transfer from point cloud sequences to meshes of diverse articulated characters. 
\end{itemize}

%% file: chapters/related_work.tex
We briefly review here relevant lines of work in automatic rigging and structure understanding from point cloud motion. 

\paragraph{Template-based rigging.} In the pioneering work of Pinocchio, Baran \etal~\shortcite{Baran:2007:ARA} proposed fitting  input skeleton templates to a given 3D mesh through an optimization procedure. Each template has specific topology, fixed number of joints, and is suited for a
specific class of characters.
Instead of hand-tuned optimization, Li \etal~\shortcite{neuralblendshape} proposed a neural network approach to estimate joint positions for a given skeletal template. They also improve skeleton-based deformations through a neural blend shape technique.
An advantage of such template-based approaches is that motion capture data for a particular template
can easily be applied. However, they are only able to rig characters whose articulation structure is compatible with the given template. A different template and network has to be trained for characters with different articulation structure (e.g., different number of parts or bone connectivity).  

\paragraph{Template-free rigging.} Another family of approaches attempts to discover the underlying articulation structure of a given shape without relying on templates. Geometric methods create skeletons based on the medial axis or centerline of shapes
\cite{Au08,Tagliasacchi12,huang13l1medial}.
However, these methods often fail to discover underlying moving parts since they rely entirely on shape geometry.  
Neural networks approaches have also been proposed to infer the articulation structure from the input geometry of a shape \cite{Xu19skeleton,rignet}.
 These networks are trained on large databases of rigged characters, thus they tend to extract skeletons that often correspond well to articulated parts. Despite their advantage of handling  characters with diverse articulation structures, these methods operate on  single static shapes without considering any motion cues. As a result, their resulting skeletons might be incompatible with  desired animations. 
Our approach is inspired by such learning-based methods. It also re-uses some of the mesh neural networks proposed in  \cite{rignet} for joint and skinning prediction. 
 Yet, the main difference is that our method 
considers both geometry and motion cues as 
input for rigging. 
Furthermore, our method is able to animate the resulting rigs automatically given the motion captured in an input point cloud sequence. 

\paragraph{Skinning.}
Another aspect of rigging is skinning  prediction, which defines the influence from each bone to each vertex.  Various geometric-based methods have been proposed for skinning ~\cite{Kavan07,Binh16, Dionne13, Jacobson11}. Yet these often require hand-tuning. Liu \etal~\shortcite{Liu2019} takes as input a skeleton template and predicts skinning  based on a learned graph neural network. Our method also learns to predict skinning, yet the predictions rely not only on geometric, but also motion cues.

\paragraph{Rigging from mesh sequences.} Several methods have been proposed to predict skeletons, including skinning weights, from sequences of meshes \cite{skinmeshanim, Kavan10CGF, Le12TOG, De08CGF, le2014robust}. However, these mesh sequences must be manually created through modeling  operations. They must be complete, noiseless, and have one-to-one correspondences between all their vertices. Instead of relying on manual 
effort to create  mesh sequences, our method
performs rigging driven by single-view performance capture of characters through  commodity sensors.   

\paragraph{Structure understanding from point clouds.}
Several methods have been proposed to discover articulation structure from point clouds 
\cite{li2007projective,Coseg16,li2016mobility,yi2018deeppart,xu20193d,Wang_2019_CVPR,RPMNet19,hayden2020nonparametric}. Given multiple observations of a captured object, they segment each point cloud into underlying moving parts. They tend not to produce consistent parts across the point cloud sequence, thus often require some form of synchronization
\cite{huang2021multibodysync}. Our goal is different in the sense that we aim to extract an animation skeleton that is consistent with both the input point cloud sequence and a target mesh. Another key  technical difference is our use of a motion transformer to encode vertex trajectories.
A more similar line of work to ours attempts to fit skeletons or deformation graphs to point cloud sequences \cite{zhang:neucom:2013,Tzionas:ECCVw:2016,lu2018unsupervised,lu20193d,bozic2021neuraldeformationgraphs}. However, their outputs are specific to the input point clouds, and cannot readily be transferred to meshes or to other characters with different part proportions. 
Our method instead takes into account both the geometry of a target mesh and the motion in the point cloud sequence for character rigging.

%% file: chapters/method.tex
\begin{figure*}[t!]
  \centering
  \includegraphics[width=1\linewidth]{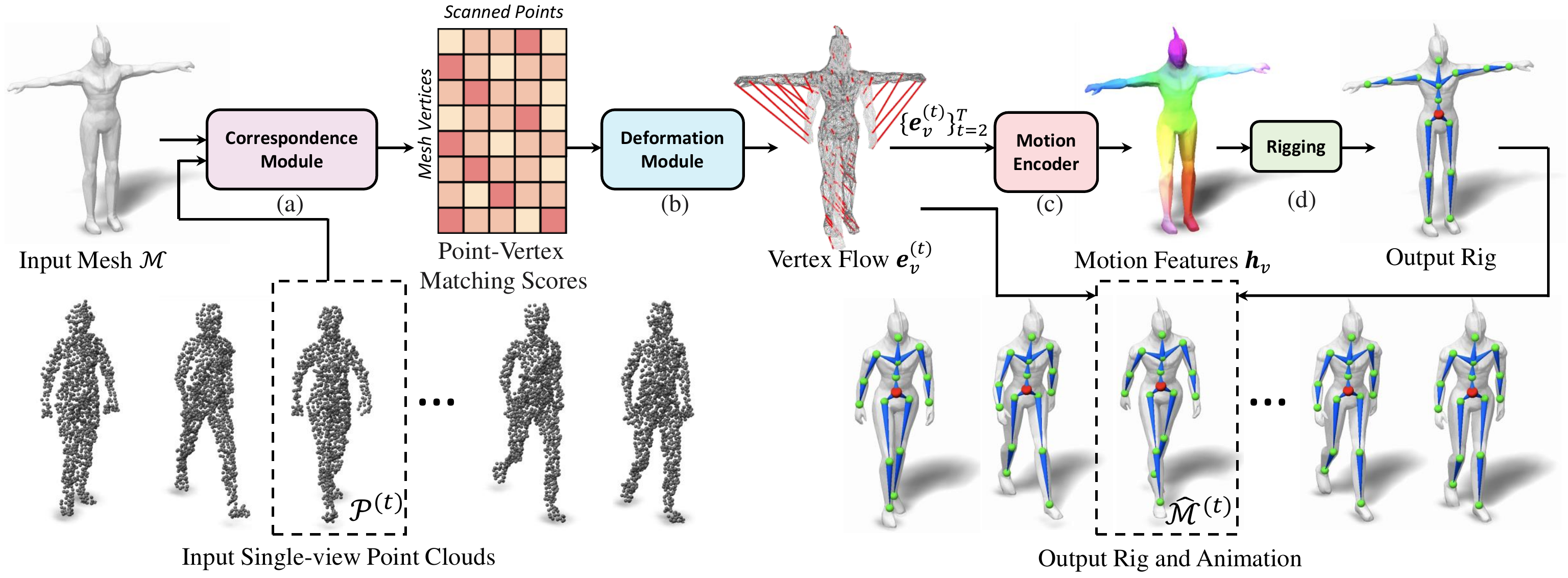}
  \vspace{-7mm}  
  \caption{Pipeline of our method: (a) The correspondence module predicts partial correspondences between the mesh and each point cloud. 
  (b) The deformation module aligns the target mesh with each point cloud frame driven by these correspondences while being robust to occlusions. (c) The motion encoder converts the resulting per-vertex  trajectories to motion-aware features that are correlated with underlying articulated parts. (d) The rigging module outputs a character rig by utilizing these features. The input mesh is animated according to the rig and the point cloud motion.}  
\label{fig:architecture}
\vspace{-3mm}  
\end{figure*}

\paragraph{Overview.} Given an input \emph{target} mesh $\mM$ representing an articulated character, the goal of our method is to rig it such that it can be animated according to the motion captured in a sequence of input point clouds  $\mP=\{\mP^{(t)}\}_{t=1}^T$, where $T$ is the number of frames in the sequence. We assume that these point cloud frames capture different poses of a \emph{reference} character whose moving parts correspond to the underlying articulated parts in the target mesh $\mM$. For example,  if the animator expects the target skeletal rig to contain bones for articulated parts, such as arms, legs, head, ears, and so on, these parts should be (a) present in the captured reference character, (b)  captured under different poses in the input  point clouds. Our method is designed to handle both partial observations, e.g., single-view point clouds with occlusion, and also mismatches in part proportions e.g., different limb length and diameter between the reference and target characters. We also note  that we do not assume any prior knowledge about the number or type of parts in the target character -- our goal is to discover them driven by the point cloud motion of the reference character. After rigging, our method can use the rig to 
produce an animated mesh sequence $\hat{\mM}=\{\hat{\mM}^{(t)}\}_{t=1}^T$
that matches the motion of the input point clouds.

The pipeline of our method is highlighted in Figure \ref{fig:architecture}. The first stage of our pipeline is a \emph{correspondence module}. Its goal is to (a) assign the target mesh vertices with a \emph{soft correspondence mask} representing the probability for them to have a corresponding point per frame, and (b) a \emph{matching score} between each vertex and each point per frame.
Its outputs drive a \emph{deformation module}
such that (a)\ vertices assigned with high probability in the correspondence mask are deformed towards matching points (b) the deformation is also propagated to vertices that have low probability in the mask due to occlusions or any small structural differences between the target and reference character. As a result, the deformation module closely aligns the target mesh with each point cloud frame. 
 
 The deformed meshes help establish per-vertex motion trajectories, which are subsequently processed by  a \emph{motion encoder} module. The module has the form of a transformer encoder that aims to convert the complete motion trajectories across all frames into motion-aware features per each vertex on the target mesh. The features are highly correlated with underlying parts i.e., vertices belonging to different articulated part tend to have distinct features. They are therefore particularly informative for locating skeletal joints and assessing the skinning weights on the target mesh (Figure \ref{fig:motion2rig}). The skeleton and skinning predictions are performed through a \emph{rigging} module, which follows RigNet \cite{rignet}, yet with the difference that this module utilizes as input our motion-aware features. The output rig is used to produce an improved mesh deformation sequence which is consistent with both the inferred skeleton and the point cloud sequence.
In the rest of this section, we describe the above modules \emph{at test time}. Training is discussed in Section \ref{sec:training}.    

\vspace{-1mm}
\subsection{Correspondence Module} This module computes partial correspondences 
between the vertices in the target mesh $\mM$ and the points per point cloud frame $\mP^{(t)}$. 

\paragraph{Point feature extraction.} To perform this task, we process the point cloud in each frame $\mP^{(t)}$ through a PointNet++ network  \cite{qi2017pointnet++} to output a feature descriptor $\bff_p^{(t)}$ for each point $p^{(t)}$ at frame $t$. The output feature is normalized to unit length (details on our PointNet++ architecture variant are included in the supplement). 

\paragraph{Vertex feature extraction.} In addition, for each target mesh vertex $v \in \mM$, we extract a feature descriptor $\bff_{v}$ through a GMEdgeNet network performing graph convolution on the target mesh
\cite{rignet}. The network relies on a message passing procedure, in which each mesh vertex accumulates messages from both its topological (one-ring) and geodesic neighbors on the mesh. 
By stacking several layers of message passing,
each vertex encodes information from a  larger context around it. The output per-vertex feature descriptor $\bff_{v}$  from the last layer is also normalized to unit length
 (architecture details can be found in the supplement).    

\paragraph{Point-Vertex matching scores.} The  matching score between each mesh vertex and input point is computed through cosine similarity:  $s_{p,v}^{(t)}=\bff_{p}^{(t)} \cdot \bff_{v}$.
The networks for feature extraction are trained with supervisory signal that encourages corresponding mesh vertices and points to have high similarity, as explained in Section \ref{sec:training}.

\paragraph{Correspondence mask.} Some vertices may not have any corresponding points due to structural differences between the target and reference character, and also due to occlusions in the reference point clouds. We employ another network, in the form of a two-layer MLP, to assess the probability $q_{v}^{(t)}$ for a mesh vertex $v$ to have correspondence with any point in the point cloud frame $\mP^{(t)}$. More specifically, for each mesh vertex, the MLP\ takes as input a vector concatenating its own feature descriptor $\bff_{v}$, 
the feature descriptor $\bff_{p(v)}^{(t)}$ of the most similar point $p(v)$ at frame $t$, where $p(v)=\argmax_p ( s_{p,v}^{(t)})$,
as well as their matching score $s_{p,v}^{(t)}$. 
The MLP outputs mask probabilities with higher values for vertices with correspondences in the captured point cloud.
We found that using both these vertex and nearest point feature vectors 
helped assess the above probability 
(see our supplement for mask visualizations).

\vspace{-1mm}
\subsection{Deformation Module} Given the vertex-point matching scores and the probabilistic correspondence mask, the deformation module deforms the target mesh $\mM$ towards each frame $\mP^{(t)}$. One possibility to perform such deformation is to move each mesh vertex towards its most similar point in the frame i.e., given a vertex position $\by_v$ and its closest matching point position $\by_{p(v)}^{(t)}$ at frame $t$, the vertex displacement could be computed as $\bd_v^{(t)}=\by_{p(v)}^{(t)} - \by_v$. This strategy is far from optimal since the displacements can easily be affected by  noise, occlusions, and overall uncertainty. We found that a better strategy is to  displace each vertex towards a weighted average of points using the matching scores as weights, i.e., $\bd_{v}^{(t)}= (\sum\limits_{p} \exp( s_{p,v}^{(t)} / \tau)\ \cdot (\by_{p}^{(t)} - \by_v)) / \sum\limits_{p} \exp( s_{p,v}^{(t)} / \tau ) $ 
where $\tau$ is learned.

\paragraph{Deformation propagation.} Still, the above displacements become unreliable for vertices with low probability $q_{v}^{(t)}$ of having correspondences. We deem displacements as unreliable for vertices whose correspondence probability $q_{v}^{(t)} < 50\%$ (the threshold is set empirically). For these vertices, we replace their displacements with the ones coming from their closest and reliable geodesic neighbor (i.e., with  $q_{v}^{(t)} \geq 50\%$).  
This encourages these vertices with no correspondences to deform more coherently with the rest of the shape. However, the deformation  tends to remain discontinuous in some areas since 
the displacement is separately calculated for each vertex and also due to the sensitivity of the above probability threshold. We can further improve the above displacements by processing them through another GMEdgeNet. The network here takes as input concatenated per-vertex positions,  displacements, and correspondence probabilities i.e., $\bx_v= [ \by_v, \bd_{v}^{(t)},  q_{v}^{(t)} ]$. The network performs feature transformations and message passing such that the input features are diffused 
and propagated over the mesh. The network outputs updated per-vertex displacements $\be_{v}^{(t)}$ representing more coherent deformations aligning the target mesh closely to each point cloud frame. We observed that all the above features are useful for estimating accurate deformations i.e., vertices with low correspondence probability have larger changes since their initial displacement is not reliable. 

\paragraph{Iterative deformation.} The deformation proceeds iteratively. After computing the displacements for the first frame, the mesh is deformed by applying the per-vertex displacements: $\by_v^{(1)} = \by_v +\be_{v}^{(1)}$. At the next frame, we compute correspondences  and matching scores between the deformed mesh and the next point cloud frame $\mP^{(2)}$. We then compute displacements to deform this mesh towards the next point cloud frame, resulting in updated vertices  $\by_v^{(2)} = \by_v^{(1)} +\be_{v}^{(2)}$. This  procedure iterates until the last frame. 

\vspace{-2mm}
\subsection{Motion encoder} 
\label{subsection:motion_encoder}
The motion encoder aims to transform the per-vertex estimated  displacements into motion-aware features, which help to estimate the rig. The input to the motion encoder is a per-vertex sequence of displacements $\{ \be_v^{(t)}\}_{t=2}^T$. It is important to note here that the encoding considers displacements computed after the initial deformation. The initial  displacements $\be_{v}^{(1)}$  are used to align the target mesh with the first poind cloud frame, and accounts for both pose and shape differences between the target and reference character.\ The rest of the displacements represent motion trajectories relative to the initially deformed mesh. We extract motion-aware features representing the motion
of this  deformed mesh, which also drive the rigging process. Since the features are computed with respect to this deformed mesh after initial displacement, the  skeletal extraction and skinning (described in Sec.~\ref{sec:rigging}) are  performed in this deformed mesh, and then transferred to the original target mesh.

\begin{figure}[t!]
    \centering
    \vspace{-2mm}
    \includegraphics[width=0.85\columnwidth]{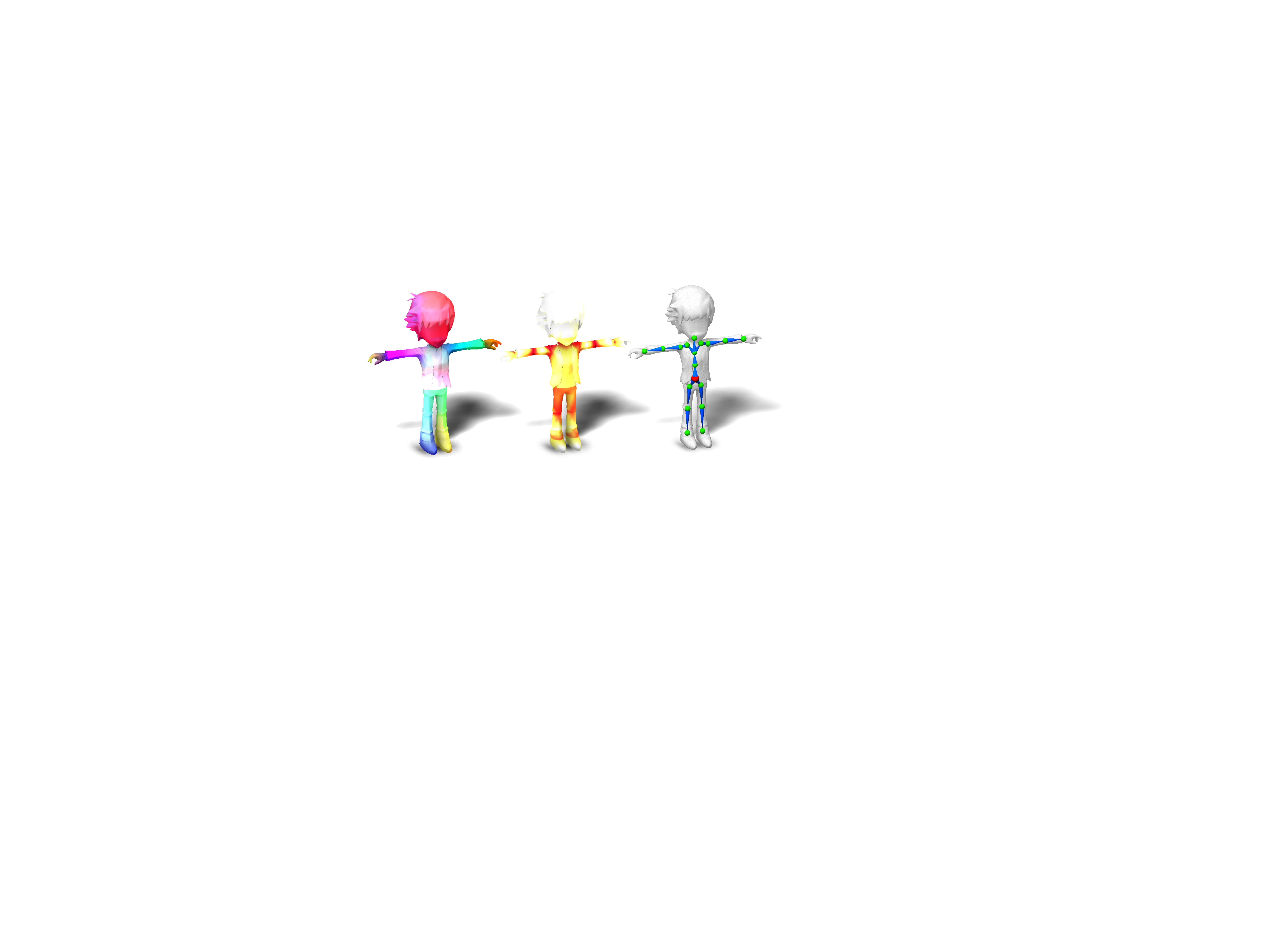}
    \vspace{-4mm}
    \caption{Our motion features (left) act as useful indicators for finding vertices close to candidate joints. Vertices with assessed high relevance to joints are shown in redder color
    in the middle. These  vertices are clustered by the rigging module to
    discover joints 
     in the animation skeleton (right).}
    \label{fig:motion2rig}
    \vspace{-4mm}
\end{figure}

\paragraph{Transformer encoder.} We tried a number of strategies to encode the per-vertex trajectories. We found that the best performance was achieved through a \textit{transformer-based} encoder module. Specifically, for each frame, the encoder first embeds each vertex displacement to a $D$-dimensional feature vector through a GMEdgeNet ($D=32$ in our implementation). This results in a
 sequence of features  $\{ \bg_v^{(t)}\}_{t=2}^T$.  We insert an extra  token with a learned embedding $\bg_v^{(1)}$ at the beginning of the sequence via another MLP, similar to the \textit{[CLS]} token in vision transformers \cite{dosovitskiy2021an}. This token plays the role of representing the whole sequence (and not the first frame). The transformer has multi-attention heads, each of which transforms the above features into query, key, value representations following \cite{vaswani2017attention}:
\mbox{$
\bq_{v,n}^{(t)} = \bQ_n \bg_v^{(t)}, \,\,\,\,
\bk_{v,n}^{(t)} = \bK_n \bg_v^{(t)}, \,\,\,\,
\bv_{v,n}^{(t)} = \bV_n \bg_v^{(t)}
$},
where $n$ denotes the attention
head index, and $\bQ_n$, $\bK_n$, $\bV_n$ are learnable parameters for query, key, and value transformations respectively. These parameters are shared across all mesh vertices. We then compute
attention matrices: $\bA_{v,n}^{(t)} = softmax( \bq_{v,n}^{(t)} \cdot \bk_{v,n}^{(t)}/sqrt(D))$. 
The attention matrices are used to compute the per-vertex motion-aware features encoding the whole sequence:
${\bg'}_{v,n}^{(1)}=\sum_{t=1}^T \bA_{v,n}^{(t)} \bv_{v,n}^{(t)}$. The features from all attention heads are concatenated,  are projected back to a $D$-dimensional feature vector with a linear transformation $\bU$ shared across all vertices: 
\mbox{$\bh_{v}= [ {\bg'}_{v,1}^{(1)},...,{\bg'}_{v,N}^{(1)} ] \bU$},  where $N$ is the number of attention heads. The  feature vector $\bh_{v}$ represents
the per-vertex motion-aware features encoding its trajectory. Details about the motion encoder architecture are provided in the supplement.

\vspace{-1mm}
\subsection{Rigging}
\label{sec:rigging}
The per-vertex motion features $\bh_v$ are used by RigNet modules \cite{rignet} for joint extraction and skinning. Both joint and skinning are estimated on the mesh after the initial deformation step. Below, we discuss these modules along with our changes.

\paragraph{Joint Extraction.} 
The joint extraction in RigNet is performed through a combination of regression and  clustering. In the regression step, the mesh vertices are displaced towards  nearest candidate joint locations. 
Since not all vertices are equally useful to determine joint locations, 
another GMEdgeNet assesses a per-vertex scalar value representing the confidence of localizing a joint from that vertex. After displacement, the vertices are clustered into joints. 
Our motion-aware features are particularly useful to better assess the importance of vertices: by computing the difference between the motion features of a vertex with respect to each of its neighbors, then aggregating these differences through pooling (as  done in GMEdgeNet), we can localize boundaries between articulated parts that are close to joint areas (Figure \ref{fig:motion2rig}, middle). Vertex displacements can also benefit from  motion-aware features since they  are highly similar within parts, making the displacements and resulting clustering into joints more coherent  (Figure~\ref{fig:motion2rig}, right). Thus, for both GMEdgeNets used for joint displacement and the vertex relevance map, we take vertex positions with their motion-aware features 
as input. The resulting joint locations are significantly improved compared to the original RigNet that uses only vertex positions as input (Figure \ref{fig:rig_comp}).
After joint discovery, 
we use RigNet's minimum spanning tree procedure to form the animation skeleton.
If the target mesh has bilateral symmetry,
we reflect discovered joints according to this mesh symmetry plane to enforce symmetric skeletons. 

\paragraph{Skinning.}\ For skinning prediction, RigNet  uses a GMEdgeNet processing vertex positions as well as volumetric distances to nearest bones. Similarly, we append the motion-aware features as additional input to the network. As a result, the skinning weights become more coherent within articulated parts since the motion-aware features tend to be similar across their vertices (Figure~\ref{fig:skinning_comp}). 

\paragraph{Rig transfer.} Our next step is to transfer the skeleton and skinning weights from the deformed mesh, whose vertex positions are $\by^{(1)} =\{\by_v^{(1)}\}_{v=1}^V$, to the original target mesh with vertex positions $\by = \{\by_v\}_{v=1}^V$, where $V$ is the number of vertices. These two meshes have explicit correspondences between their vertices, yet they  may differ in  pose and part proportions. 
There are various strategies to transfer skeletons across meshes driven by vertex correspondences \cite{allen03,Avril16}, however they 
use hand-engineered energy optimization approaches or are sensitive to select particular vertices as markers. In our case, we resorted to fit a joint regressor that maps the 3D positions of vertices to joint locations: $\bj = f(\by)$, where $\bj=\{\bj_i\}_{i=1}^J$ are joint positions, $J$ is the number of joints extracted from the rigging step, and $f$ is a neural network interpolating function in the form of a MLP with a single hidden layer. Fitting is performed with gradient descent using as training data variants of the mesh $\by^{(1)}$. The variants are created by applying random rotations and anisotropic scaling transformations to the joints of that mesh. We found that sampling about $300$ variants promotes an accurate fitting of joint positions under different mesh configurations i.e., the average Chamfer distance between predicted joints and ground-truth ones in neutral-pose meshes (unseen during training) was less than $0.1\%$ of the largest  bounding box axis of test meshes in the ``ModelsResource'' dataset (discussed in our experiments). The fitting is fast, taking $3$ sec per mesh (measured on a NVidia 2080Ti).
After fitting, we estimate the joint positions with the MLP.
The skinning weights are copied to corresponding vertices from the deformed mesh to the target one.

\paragraph{Motion transfer.} Given the motion trajectories of the deformed mesh vertices $\by^{(1)}$ and the skeleton, we apply full-body IK ~\cite{parent2012computer}
 to compute joint angles  such that the deformed mesh follows the point cloud motion. The joint angles are transferred to the original mesh through IK-based retargeting \cite{Lee99}.

%% file: chapters/training.tex
\begin{figure*}[!t]
     \centering
     \includegraphics[width=0.95\textwidth]{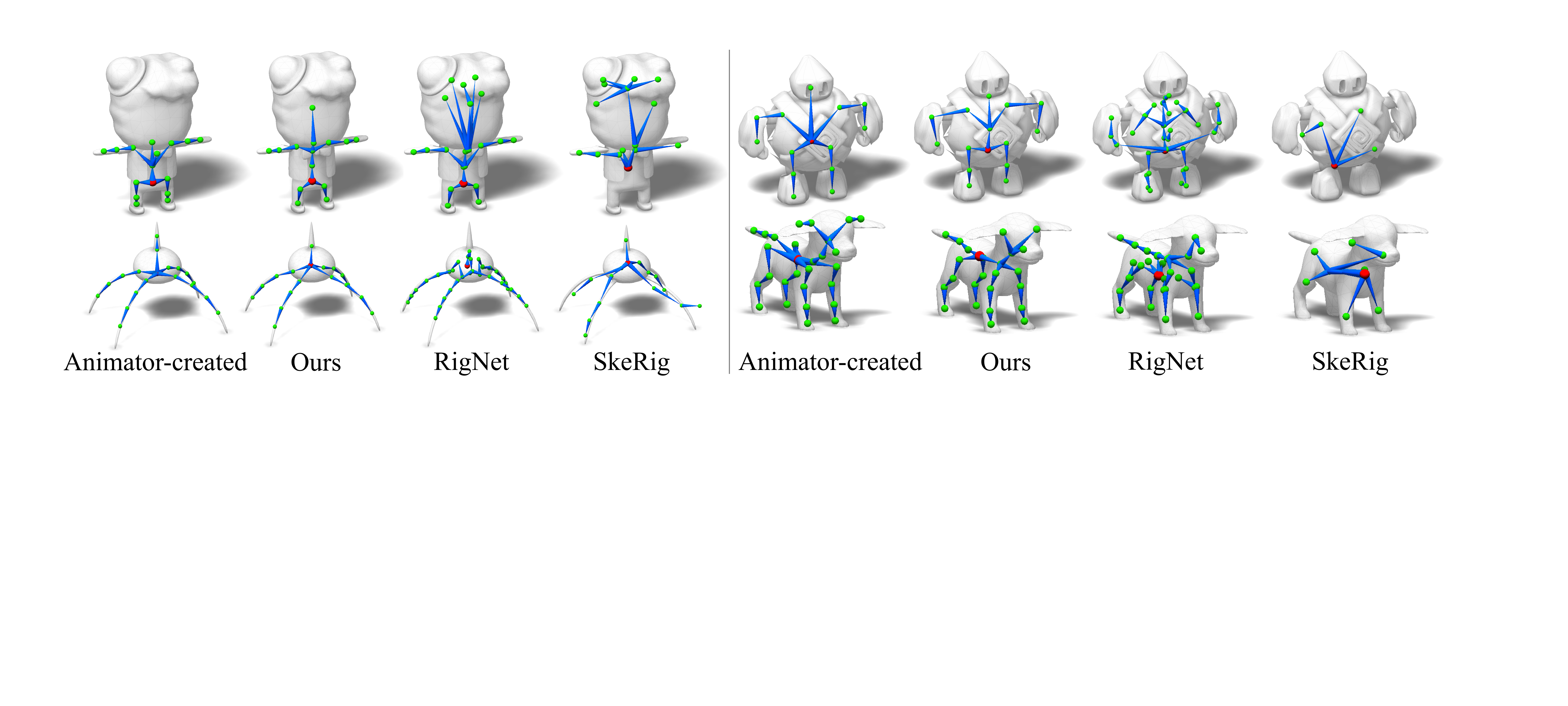}
     \vspace{-4mm}
     \caption{Comparisons with previous methods for skeleton prediction. For each character, an animator-created skeleton is shown on the left as reference. With the help of motion information, our prediction captures articulated parts more accurately resulting in skeletons that agree more with the artist-made ones.}
     \label{fig:rig_comp}
     \vspace{-0.5em}
\end{figure*}

\begin{figure}[!t]
     \centering
     \includegraphics[width=0.95\columnwidth]{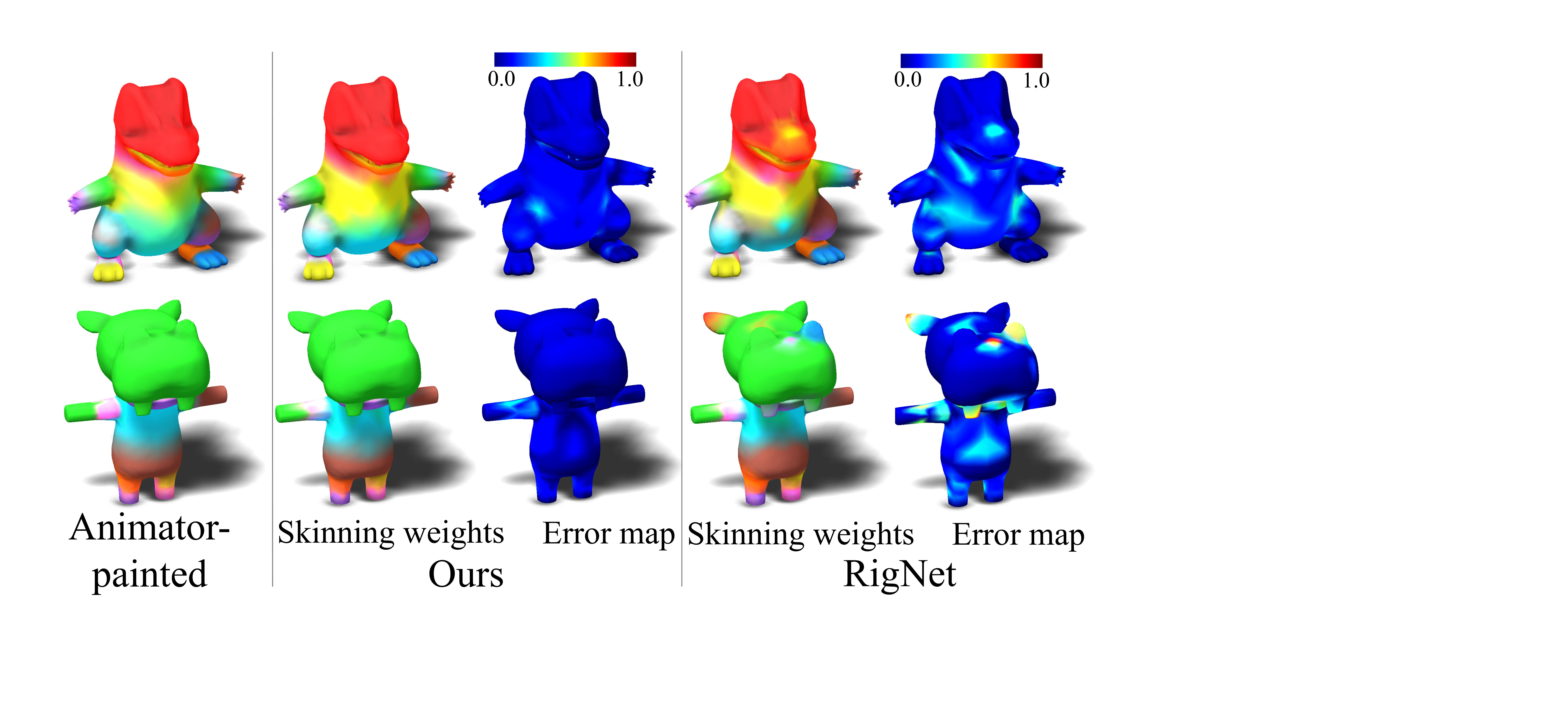}
     \vspace{-3mm}
     \caption{Comparison with RigNet on skinning prediction. To evaluate skinning alone, we use the animator's skeleton as input to the skinning module of both methods. We show (a) the animator's skinning weights, (b,c) MoRig's skinning and error map, and (d,e) RigNet's skinning and errors.\\  }
     \label{fig:skinning_comp}
     \vspace{-7mm}
\end{figure}

We now describe the training of all the modules in our pipeline.

\paragraph{Datasets.}
We rely on two datasets for training our modules.
First, we make use of the training split of the ModelsResources dataset~\cite{rignet}, which contains $2,163$ rigged meshes.
Since there are no animations associated with these characters, we synthesize animation sequences for each mesh by adding random rotations to the joints represented in keyframes, then interpolate the motion resulting in a $100$-frame sequence per character (total $21,630$ frames). 

As a second dataset, we use the DeformingThing4D dataset~\cite{li20214dcomplete}. The dataset has the advantage of containing realistic motion data for humanoid characters based on motion capture, as well as animations for  non-humanoid characters created by artists. For training, we use $83$ characters from this dataset. Each model has multiple animation sequences (on average, $15.44$ sequences per character). The total number of frames is $128K$. We note that the dataset does not contain any animation skeletons or skinning. 

Both datasets do not contain any associated point clouds. To this end, for each of their animation sequences, we render depth images for each frame by randomly placing a camera in the frontal viewing hemisphere of the character and sampling random azimuths and elevations between $[-36,+36]$ degrees. We back-project pixels from depth to 3D space via the camera parameters. To better simulate scans, we add Gaussian noise to the depth map with $\sigma=0.05$. Even if the point clouds are generated synthetically, we observed sufficient generalization to several real-world motion capture characters.

\paragraph{Correspondence module training.}
To train the correspondence module, we apply supervisory signal for both
matching scores and correspondence mask. Since for each training mesh, the point cloud is generated synthetically, we have ground-truth (a) visibility masks $m^{(t)}_v$ representing whether a vertex $v$ is visible in the partial point cloud at frame $t$, (b) ground-truth correspondences between mesh vertices and their closest points in the point cloud per frame. We use binary cross entropy to supervise correspondence probability learning:
$
L_{mask} = \sum_{t,v}  BCE( q^{(t)}_v, m^{(t)}_v)
\label{eq:vis_loss}
$.
We also use the InfoNCE loss~\cite{oord2018representation} to favor the feature of each visible mesh vertex to match the one of its corresponding point:
\begin{equation}
L_{corr} = 
    -\sum_{t,v}\log
    \frac
    {\exp{(\bff_{p(v)}^{(t)} \cdot \bff_{v} /\tau)}}
    {\sum_{p' \ne p(v)} {\exp{(\bff_{p'}^{(t)} \cdot \bff_{v} /\tau)}}}
\label{eq:corr_1_loss}
\nonumber
\end{equation}
where  $\tau$ represents a learnable temperature
~\cite{radford2021learning}. Similarly, we use the same loss to match the representation of each point with its corresponding vertex.

\paragraph{Deformation module training.}
\label{sec:flow_training}
To train the GMEdgeNet used in the deformation module, we provide supervisory signal in the form of ground-truth displacements $\hat\bd_{v}^{(t)}$
for each mesh vertex $v$ at frame $t$ in the training animation sequences. We use the L1 loss to penalize deviation of predicted displacements from the ground-truth ones:
$
\mbox{$L_{flow} = \sum_{v,t} | \hat\bd_{v}^{(t)} - \bd_{v}^{(t)}|_1$}
\label{eq:flow_loss}
$.
We pre-train the correspondence and the deformation modules on the synthetic sequences of the ModelsResource dataset, then fine-tune them on the sequences provided in the DeformingThing4D dataset.

\paragraph{Motion encoder training.} To train the transformer encoder, we provide supervisory signal such that the output motion features $\bh_{v}$ are similar for vertices with similar rigid motion. The training skinning weights can help us assess such vertices. Specifically, if two vertices of the mesh have almost identical skinning weights, they are expected to have similar motion features. Given two vertices $v$ and $u$, we examine if the L1 difference   of the ground-truth skinning weights $|\bs_v - \bs_{u}|$ is small (we use a threshold $0.1$). We use the InfoNCE loss to promote similar motion features for such vertices: 
\begin{equation}
L_{motion} = 
    -\sum_{v,u}\log
    \frac
    {\exp{(\bh_v \cdot \bh_{u} /\tau')}}
    {\sum\limits_{v' \in \mR(v)} \exp{(\bh_v \cdot \bh_{v'} /\tau')}
    }
\label{eq:motion_loss}
\nonumber
\end{equation}
where $\mR(v)$ is the set of vertices whose skinning weights differ from vertex $v$ based on the above threshold, and $\tau'$ is a learnable parameter. We note that we use this loss on the  ModelsResources training split only, which provides  ground-truth skinning weights.

\paragraph{Rigging module training.} To train the rigging module, we follow the same losses under the same training procedure described in \cite{rignet} on the  ModelsResources training split.

%% file: chapters/experiments.tex
\begin{figure*}[!t]
     \centering
     \includegraphics[width=1\textwidth]{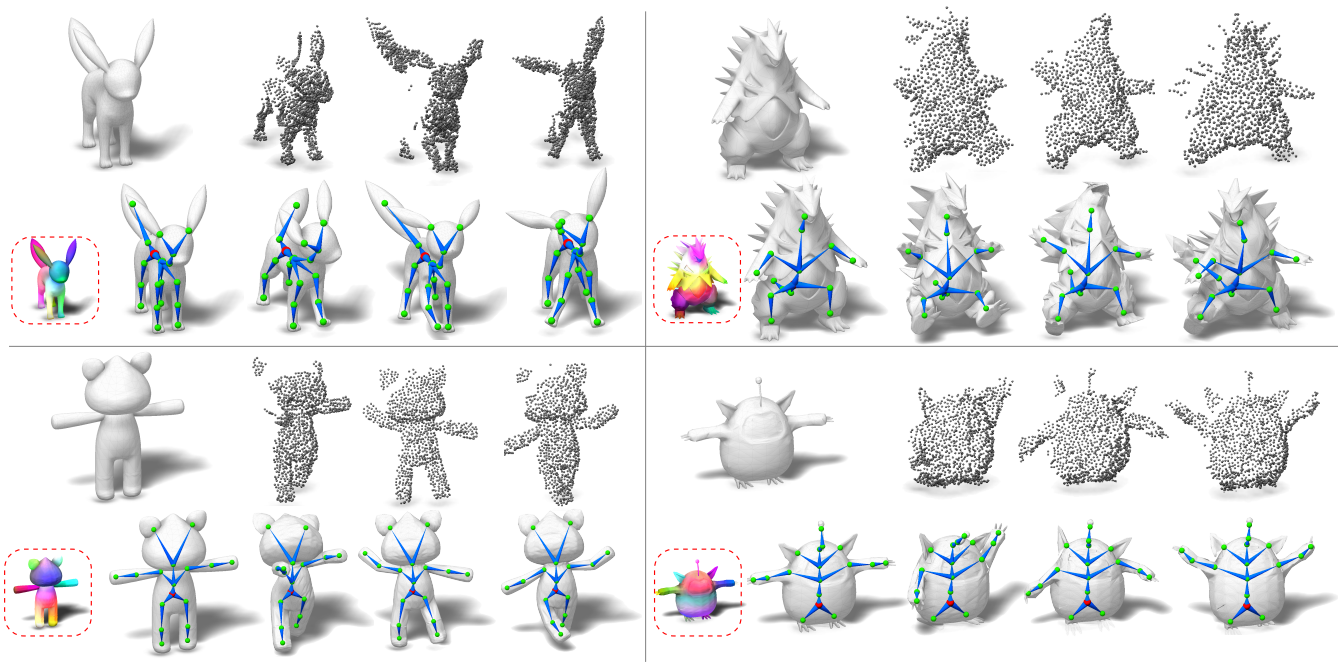}
     \vspace{-7mm}
     \caption{Rigging and animation results from  synthetic point cloud sequences of the ModelsResource dataset. For each example, we show the target mesh and representative point cloud frames from the input sequence. We also show our  motion-aware features (red rectangle), along with the resulting rigs and deformed meshes corresponding to the point cloud frames.}
     \vspace{-2mm}
     \label{fig:modelsresource}
\end{figure*}

\begin{table}[t!]
\small
\centering
\begin{tabular}{|@{}c@{}|@{}c@{}c@{}c@{}c@{}|@{}c@{}c@{}c@{}|@{}c@{}|}
\hline
& \multicolumn{4}{c|}{Joint evaluation}   & \multicolumn{3}{c|}{Skinning evaluation}            & Anim.       \\
\hline
& \,Chamfer\, & IoU             & \,Precision\,           & \,Recall\,            & \,Precision\,           & \,Recall\,            & \,Avg L1\,        & \,L2 dist\, \\
\hline
SkeRig  & 7.5\%          & 26.5\%          & 33.4\%          & 23.2\%          & N/A             & N/A             & N/A           & 2.9\%\           \\
RigNet                  & 3.9\%          & 61.6\%          & 67.6\%          & 58.9\%          & 82.3\%          & 80.8\%          & 0.39          & 3.1\%\           \\
MoRig                & \textbf{3.5\%} & \textbf{64.7\%} & \textbf{72.2\%} & \textbf{61.2\%} & \textbf{83.7\%} & \textbf{85.5\%} & \textbf{0.32} & \textbf{2.4\%} \\
\hline
\hline
avg pool   & 3.7\%         & 61.8\% & 67.2\% & 56.0\% & 77.5\%      & 78.6\%      & 0.35        & 2.9\%           \\
max pool       & 3.6\%         & 63.2\% & 70.5\% & 58.8\% & 81.4\%      & 83.2\%      & 0.34        & 2.8\%           \\
\hline     
\end{tabular}
\caption{ Comparison with other methods and MoRig variants.}
\vspace{-7mm}
\label{table:comparison} 
\end{table}

In this section, we describe our experiments.
Our code and data are provided at \textcolor{blue}{
\href
{https://github.com/zhan-xu/MoRig}
{https://github.com/zhan-xu/MoRig}}.
Please see our video for animations and more results: \textcolor{blue}{\href{https://youtu.be/sPxfnQ8j07Y}{https://youtu.be/sPxfnQ8j07Y}}.

\paragraph{Comparisons.} Since our method does not rely on particular specific skeleton templates, we compare with other template-free rigging methods.
We first compare with RigNet \cite{rignet}. RigNet does not consider any motion cues as input. 
The main point of this comparison is to show that  motion cues are useful to achieve better rigging. We note that RigNet is trained on the same split as our method,  uses the same training losses for its rigging module, and its hyperparameters are tuned   in the same validation split as ours. Second, we compare with \cite{le2014robust} (abbreviated as ``SkeRig''), 
an  optimization-based method for rigging based on input mesh sequences. This method does not process input point clouds. It instead processes mesh sequences to produce both skeletons and skinning for an input target mesh.
We provide this method with the target mesh and the corresponding deformed mesh sequence produced by our deformation module. The goal of this comparison is to test whether the proposed deep motion features and rigging modules are better than an optimization method that fits joints and skinning to the mesh sequence produced by our deformation stage.

\paragraph{Qualitative comparisons.} 
Figure \ref{fig:rig_comp} shows comparisons of our method with RigNet and SkeRig on test animation sequences from the test split of the ModelsResource dataset.
In all examples, the number and placement of joints is more accurate for MoRig.
Figure \ref{fig:skinning_comp} shows comparisons on skinning weights. As shown in the error maps, our method produces more accurate skinning weights.

\paragraph{Quantitative comparisons.} We use the  evaluation measures proposed in RigNet~\cite{rignet} to measure the quality of the predicted skeletons and skinning. The evaluation is performed on 
the test split  of ModelsResource, which contains $270$ characters along with artist-made rigs and our $270$ synthetically generated animation sequences (one per test character).
As seen in Table \ref{table:comparison}, our method outperforms both RigNet and SkeRig for joint prediction. For example, in terms of Chamfer distance between artist-specified and predicted joints, our method offers a relative reduction of $53\%$ compared to SkeRig ($7.5\%$ $\rightarrow$ $3.5\%$ distance error relative to the character's height), and by $10\%$ compared to RigNet ($3.9\%$ $\rightarrow$ $3.5\%$). Our supplement includes more evaluation wrt real scans -- same trends were observed there.

Regarding evaluation wrt skinning, we use the same animators' skeleton as input to competing methods such that differences in skeleton prediction are factored out. We here compare only with RigNet, which can compute skinning weights given an input skeleton. As shown in Table \ref{table:comparison}, MoRig outperforms RigNet by yielding a $18\%$ relative reduction in L1 skinning weight error ($0.39 \rightarrow 0.32$). 

We also compared the methods in terms of how well they reproduce the ground-truth animation sequence.
Specifically, given the ground-truth deformed mesh sequence on the ModelsResource test split, and the predicted ones from all three methods based on their predicted skeleton and skinning weights, we measure the Euclidean distance error
between ground-truth
vertex positions and the ones produced by each method. The error is averaged over all vertices and frames in the test split. Our method is able to produce deformed sequences much closer to the ground-truth ones (see `Anim.' column, Table \ref{table:comparison}). This highlights MoRig's compound improvement in both skeleton and skinning weight prediction. 
Compared to RigNet, we observe a relative error reduction of $22.5\%$ ($3.1\% \rightarrow 2.4\%$ distance error with respect to the height of character). Compared to SkeRig, we also see a significant error reduction $17.2\%$ ($2.9\% \rightarrow 2.4\%$).

\paragraph{More qualitative results}
Figure \ref{fig:modelsresource} shows four examples of rigging and deformation results from the ModelsResource dataset. The bottom row of Figure \ref{fig:teaser} shows an additional result of a test quadruped rigged and animated by motion from the DeformingThing4D dataset (not included in our training). For animation results, please see our supplementary video
(\textcolor{blue}{\href{https://youtu.be/sPxfnQ8j07Y}{https://youtu.be/sPxfnQ8j07Y}}).

\begin{figure*}[!t]
     \centering
     \includegraphics[width=1\textwidth]{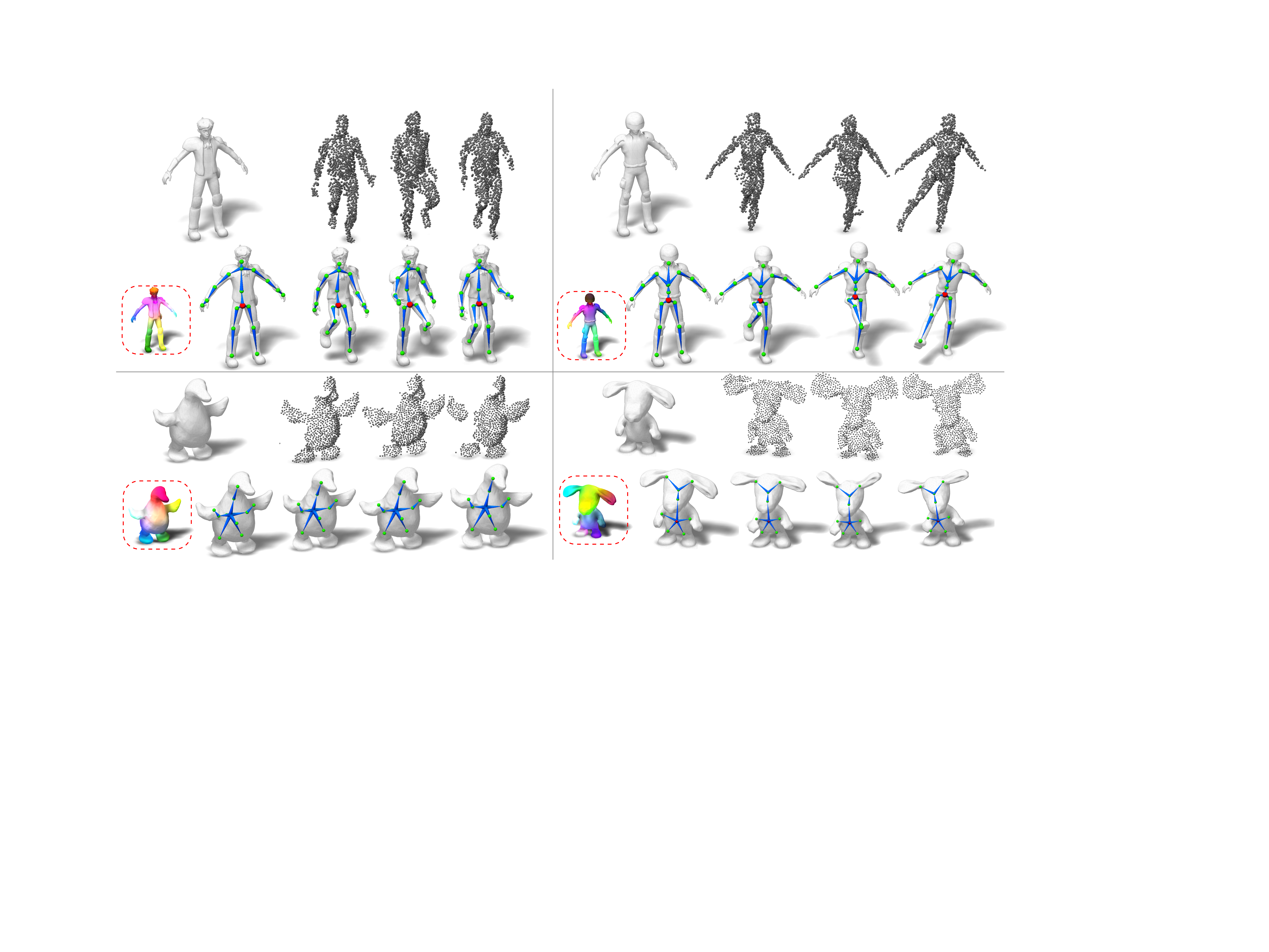}
     \vspace{-7mm}
     \caption{Rigging and animation results with real-world point cloud sequences from DFaust (top) and KillingFusion~\cite{slavcheva2017cvpr} (bottom).  
      For each example, we show the target mesh and representative point cloud frames from the input sequence. We also show our  motion-aware features (red rectangle), along with the resulting rigs and deformed meshes corresponding to the point cloud frames (see also our video for animated results: \textcolor{blue}{\href{https://youtu.be/sPxfnQ8j07Y}{https://youtu.be/sPxfnQ8j07Y}}).}
     \label{fig:real_data}
     \vspace{-2mm}
\end{figure*}

\paragraph{Real-world scans.} We also evaluated MoRig's rigging and animation performance on test cases involving real-world scans. The top row of Figure \ref{fig:teaser} shows a fictional character rigged and animated using a
real-world point cloud sequence from DFaust~\cite{dfaust:CVPR:2017}. The top row of Figure \ref{fig:real_data} shows two more real-world point cloud sequences from DFaust along with resulting rigs and animated meshes from MoRig. The bottom row of Figure \ref{fig:real_data} shows MoRig's result driven by two real-world point cloud sequences from KillingFusion~\cite{slavcheva2017cvpr}. These scans are noisy and partial, with points accessible only from a single viewpoint, making the problem challenging. 
Even if our method was not trained on real-world point clouds, our method can still create skeletons capturing the articulated parts in these characters. 
Quantitatively we evaluate the joints in our rigs compared to animator-created ones for the meshes used in these real-world test cases in Table \ref{tab:numbers_real_scan}. We also provide the results of RigNet as a comparison. The results show better joint localization compared to RigNet with trends similar 
to our evaluation on synthetic data. The animated results are shown in the video (3:05-end). 

\begin{table}[t!]
\centering
\begin{tabular}{|c|c|c|c|c|}
\hline
        & IoU &  Precision &  Recall & Chamfer  \\
\hline
RigNet  &  46.7\%  &  44.5\%    &  49.9\%   &  3.3\%   \\
MoRig   & \textbf{65.7\%}   &  \textbf{70.0\%}  & \textbf{62.2\%}   &  \textbf{3.1\%}  \\
\hline
\end{tabular}
\label{tab:numbers_real_scan}
\caption{Comparison with RigNet wrt joint localization based on real-world test cases.} 
\vspace{-9mm}
\end{table}

\paragraph{Ablation study.} \setlength{\columnsep}{8pt}
\begin{wrapfigure}{R}{0.55\linewidth}
\vspace{-5mm}
  \includegraphics[width=0.5\columnwidth]{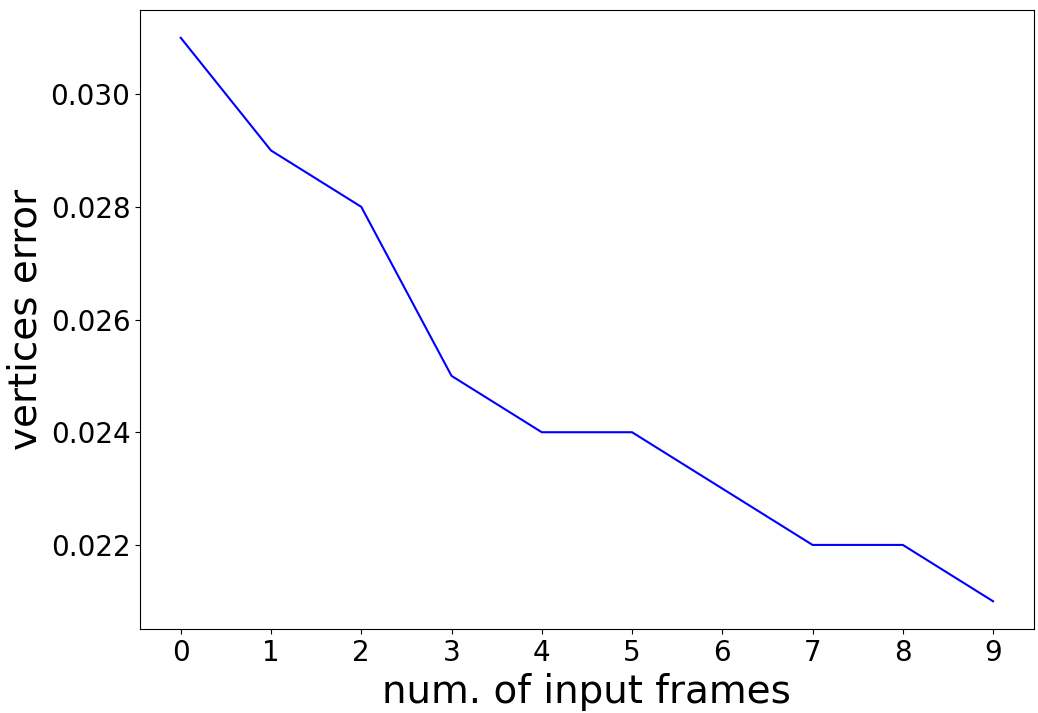}
  \vspace{-3mm}
  \caption{\small Vertex error wrt different number of input frames.}
  \vspace{-4mm}
  \label{fig:curve}
\end{wrapfigure}
We include comparisons with two alternative variants to produce motion-aware features. Instead of a transformer encoder, 
we can alternatively produce features by mean or max pooling over the features $\{ \bg_v^{(t)}\}_{t=2}^T$ (see Section~\ref{subsection:motion_encoder}). Our transformer-based encoder yields the best results, as shown in Table \ref{table:comparison}. 
We also performed another ablation study to evaluate MoRig's performance wrt the number of input frames, or poses. In Figure~\ref{fig:curve}, we show the Euclidean distance error between ground-truth vertex positions and predicted ones (as \textit{`y-axis'}) versus number of input poses (as \textit{`x-axis'}). With more frames as input to our motion encoder, the performance is improved since the features become increasingly informative about the articulated parts in the motion.

%% file: chapters/conclusion.tex
We presented MoRig, a method for rigging and animating a character mesh driven by a point cloud sequence of a captured character. 
Comparing to existing single mesh rigging approaches, our method takes as input an additional point cloud motion sequence, and encodes it into motion features via a deep neural network. The encoded motion features are informative about articulated parts of the target character and thus can achieve more accurate skeletal rigging results.

Our method has limitations that can inspire future research. We assumed that the reference and target character have similar underlying articulation structure. If a part is not visible in an input sequence (e.g., tail), our method will be entirely based on geometry instead of motion and may fail to create a bone for this part. Although we attempted to make our method robust to occlusions and mismatches between part proportions of the target and reference character, large changes in geometry and poses may result in inaccurate rigs and motion transfers. In particular, if the initial point cloud frame is corrupt, then the initial deformation becomes unreliable, and our method will fail. Our method is limited to linear blend skinning, and cannot deal with topological changes. Finally, it requires guidance in the form of point clouds, which may not always easy to capture.

%% file: chapters/suppl.tex
\begin{figure*}[!t]
\vspace{-22mm}
    \centering
    \includegraphics[height=0.48\textwidth]{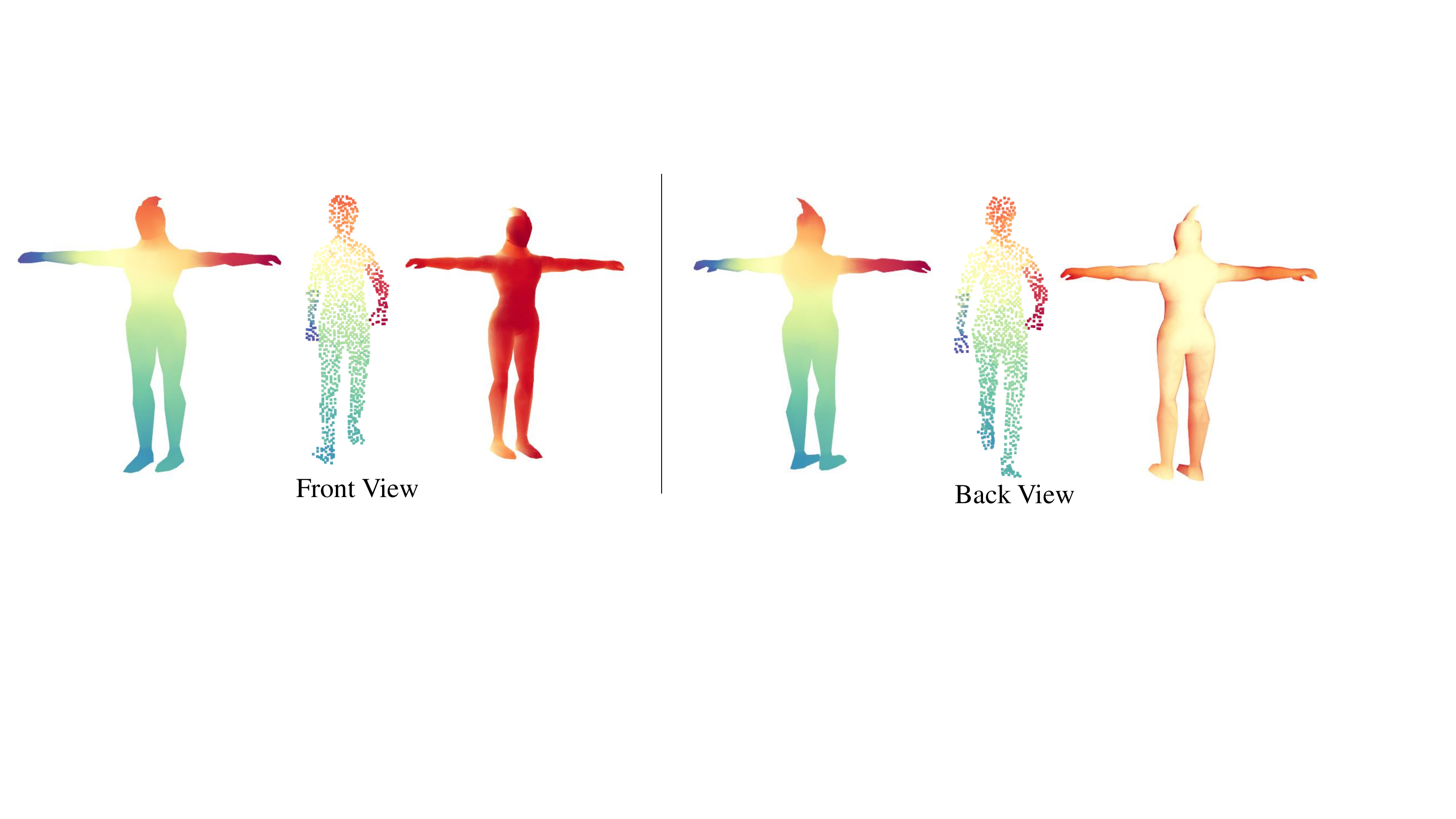}
\vspace{-36mm}
    \caption{Our predicted  correspondences between target meshes and input point clouds. We also include the correspondence mask. In each view, we show the color-coded mesh according to vertex positions (left), along with the point cloud whose color is decided by the corresponding vertex of each point (middle). We also show the correspondence mask probability map (right).}
    \vspace{0mm}
    \label{fig:corrvis}
\end{figure*}

\begin{table*}[h!]
\begin{tabular}{|c|c|c|}
\hline
Layer                                          & Input                & Output        \\
\hline
\hline
\multicolumn{3}{|c|}{Point Feature Extractor}                                         \\
\hline
SA(0.5, 0.12, {[}3, 32, 32,64{]}, 64)          & x0: (N, 3)           & (N/2, 64)     \\
SA(0.25, 0.25, {[}64+3, 64, 64, 128{]}, 64)    & (N/2, 64)            & (N/8, 128)    \\
SA(0.25, 0.5, {[}128+3,  256, 256, 256{]}, 64) & (N/8, 128)           & (N/32, 256)   \\
GlobalSA({[}256+3, 256, 256, 512{]})           & (N/32, 256)          & (1, 512)      \\
FP({[}512+256, 256, 256{]})                    & (1, 512)             & (N/32, 256)   \\
FP({[}256+128, 256, 128{]})                    & (N/32, 256)          & (N/8, 128)    \\
FP({[}128+64, 128, 64{]})                      & (N/8, 128)           & (N/2, 64)     \\
FP({[}64, 64, 64{]})                           & (N/2, 64)            & (N, 64)       \\
MLP({[}64, 64{]}) + Lin(64, 64)                & (N, 64)              & (N, 64)       \\
\hline
\hline
\multicolumn{3}{|c|}{Vertex Feature Extractor}                                          \\
\hline
GCU(3, 32)                                     & (M, 3)               & x1: (M, 32)   \\
GCU(32, 64)                                    & (M, 32)              & x2: (M, 64)   \\
GCU(64, 256)                                   & (M, 64)              & x3: (M, 256)  \\
GCU(256, 512)                                  & (M, 256)             & x4: (M, 512)  \\
MLP({[}864, 1024{]})                           & cat(x1,x2,x3,x4)     & (M, 1024)     \\
MaxPool + repeat                               & (M, 1024)            & xg: (M, 1024) \\
MLP({[}1891, 1024, 256{]}) + Lin(256, 64)      & cat(x0,x1,x2,x3,x4, xg) & (M, 64)   \\   
\hline
\end{tabular}
\caption{Correspondence module details. $N$ is the number of input points, $M$ is the number of input vertices. Note that we do not require them to be fixed. \textit{GCU} is the GMEdgeConv layer as RigNet, which encodes both geodesic and topological neighbors.} 
\label{tab:corr_module}
\vspace{-5mm}
\end{table*}

\begin{table*}[h!]
\begin{tabular}{|c|c|c|}
\hline
Layer                   & Input                    & Output        \\
\hline
GCU(7, 128)             & x0: (M, 7)             & x1: (M, 128)  \\
GCU(128, 256)           & (M, 128)                 & x2: (M, 256)  \\
GCU(256, 512)           & (M, 256)                 & x3: (M, 512)  \\
MLP({[}896, 1024{]})    & cat(x1,x2,x3)            & (M, 1024)     \\
MaxPool+repeat          & (M, 1024)                & xg: (M, 1024) \\
MLP({[}1927, 1024, 256{]})+Lin(256, 3) & cat(x0, x1,x2,x3,x4, xg) & (M, 3)        \\
\hline
\end{tabular}
\caption{Deformation module details. Symbols are defined as above. We note that GCU in this module is slightly modified to encode vertex position and additional feature with different branches.} 
\label{tab:deform_module}
\vspace{-5mm}
\end{table*}

\begin{table*}[h!]
\begin{tabular}{|c|c|c|}
\hline
Layer                                       & Input                      & Output\\
\hline
GCU(6, 64)                                  & x0: (M, 6)                 & x1: (M, 64)   \\
GCU(64, 256)                                & (M, 64)                    & x2: (M, 256)  \\
GCU(256, 512)                               & (M, 256)                   & x3: (M, 512)  \\
MLP({[}832, 1024{]})                        & cat(x1,x2,x3)              & (M, 1024)     \\
MaxPool+repeat                              & (M, 1024)                  & xg: (M, 1024) \\
MLP({[}1862, 1024, 256{]})+Lin(256, 32)     & cat(x0, x1,x2,x3,x4, xg)   & (M, 32)       \\
$\bQ_n$, $\bK_n$, $\bV_n$: Lin(32, 64)                     & (M, T, 32)                 & (M, T, 64)    \\
$\bU$: Lin(64$\times$n\_head, 64) & (M, T, 64$\times$n\_head ) & (M, T, 64)    \\
MLP({[}64, 512{]}) + Lin(512, 32)           & (M, T{[}token{]}, 64)      & (M, 32)   \\
\hline
\end{tabular}
\caption{Motion encoder details. As in the deformation module, GCU in this module is modified to encode vertex position and additional features from different branches. $T$ is the number of input frames. $T[token]$ is an index to the token feature representing the whole sequence.} 
\label{tab:motion_module}
\vspace{-5mm}
\end{table*}

\section{Implementation details}
The proposed method is implemented in PyTorch~\cite{NEURIPS2019_9015}. Training follows a stage-wise procedure. First, we apply the loss $L_{mask}$ and $L_{corr}$ in the main paper to train the correspondence module on the ModelsResource dataset using the Adam optimizer \cite{kingma:adam}. We then add the loss $L_{flow}$ to train the deformation module also on the same split. We then fine tune both modules on realistic motion sequences from the DeformThings4D dataset. Next, we train the motion encoder and rigging module together on the ModelsResource dataset with a $50\%$-$50\%$ mixture of sequences with predicted and ground-truth deformations respectively. We found that using this mixture improves the final performance as it helps the network utilize motion information while maintaining robustness to noises. In all cases, hyperparameters are tuned in the validation splits of these datasets. Training takes about $50h$ on a Nvidia RTX 8000 for all modules on both training datasets. We  provide the network architecture details of the proposed \textit{correspondence module}, \textit{deformation module} and \textit{motion encoder} in Table \ref{tab:corr_module}, \ref{tab:deform_module} and \ref{tab:motion_module} respectively along with the corresponding input and output feature sizes.

\section{Correspondence Visualizations}
Figure \ref{fig:corrvis} visualizes an example of predicted correspondences between a mesh and a point cloud, as derived from the matching scores and correspondence mask. We show the result from both front and back views. Vertices are color-coded according to their positions, and point colors are determined based on their predicted corresponding vertex. In the correspondence mask, redder color indicates the vertices are more likely to have a corresponding point.